\documentclass{article}
\usepackage{spconf,amsmath,graphicx}
\usepackage{helvet}
\usepackage{courier}
\usepackage{algorithm}
\usepackage{algorithmic}
\usepackage{amsmath}
\usepackage{float}
\usepackage{graphicx}
\usepackage{multirow}
\usepackage{subfigure}
\usepackage{changepage}

\title{Multi-Channel Auto-Encoder for Speech Emotion Recognition}
\name{Zefang Zong, Hao Li, Qi Wang}
\address{Department of Electronic Engineering, Tsinghua University, Beijing, China}

\begin{document}
\maketitle
\begin{abstract}
Inferring emotion status from users' queries plays an important role to enhance the capacity in voice dialogues applications. Even though several related works obtained satisfactory results, the performance can still be further improved. In this paper, we proposed a novel framework named multi-channel auto-encoder (MTC-AE) on emotion recognition from acoustic information. MTC-AE contains multiple local DNNs based on different low-level descriptors with different statistics functions that are partly concatenated together, by which the structure is enabled to consider both local and global features simultaneously. Experiment based on a benchmark dataset IEMOCAP shows that our method significantly outperforms the existing state-of-the-art results, achieving $64.8\%$ leave-one-speaker-out unweighted accuracy, which is $2.4\%$ higher than the best result on this dataset.

\end{abstract}
\section{Introduction}
\label{sec:intro}

As voice-controlled intelligent applications develop rapidly, emotion recognition and analysis are becoming more and more important. Obtaining information from literal expressions cannot satisfy our demand any more, for a great part of information is conveyed by human emotions. Some utterance, for example, ironic phrases, may have completely opposite meaning from what it sounds literally. And voice-controlled virtual assistants like Siri \footnote{http://www.apple.com/ios/siri/} may work much better with emotion information. So inferring emotion from voice data can help to understand the accurate meaning of users, as well as providing more humanized responses. 

Traditionally, two major frameworks were explored for speech emotion recognition. One is HMM-GMM framework based on dynamic features \cite{schuller2003hidden} , and the other one is classified by support vector machines (SVM) based on high-level representations generated by applying lots of functions on low-level descriptors (LLDs) \cite{schuller2009acoustic}.

Recently, more and more attention has been paid to deep learning methods for speech emotion recognition which brings a better performance than the traditional frameworks \cite{kim2013deep}. Some researches focus on utterance-level features which usually extract high-level representations from LLDs and then utilize deep neural network (DNN) for classification \cite{xia2017multi}. Meanwhile, instead of high-level statistics representation, some other researchers utilize frame-level representation or raw signal as inputs to neural network for an end-to-end training \cite{zhang2016multimodal,trigeorgis2016adieu}. Generally, deep learning approaches has made a great contribution to the field of speech emotion learning. 

However, the current methods using deep learning have some apparent limitations in training procedure. In the framework using utterance-level features, the features to be inputted into neural networks are always generated by concatenating LLDs directly. Not only does this method ignore the independent nature of each feature, but also results in generating too high dimensions of the input, which makes it quite hard to reach a satisfying result because of overfitting due to amount-limited training data. Although we can reduce the dimension susing other methods, the process of reduction may lose important
information, which is inevitable to an extent.

In this paper, we put forward the multi-channel auto-encoder (MTC-AE) as a new scheme to avoid the limitations listed above. Instead of training the network using all features concatenated directly, we take features from several local classifiers by utilizing DNN separately,  which can keep the information from the independence of each classifier,  and add a strong regularization to the whole system, helping to relieve overfitting. Features are taken from bottleneck layers of each classifier and then concatenated into a higher-dimension one. The final concatenated feature will be the input of a global classifier. Because of the regularization of local classifiers, the concatenated feature is more discriminative for classification. Finally, we fuse the outputs of all classifiers to obtain the final prediction results. It's important that the training procedure in both global classifier and the local ones are trained simultaneously through a single objective function, which guarantees that the final results consider both independence and relevance among different features. In addition, inspired by bottleneck features\cite{Gehring2013Extracting}, we initialize each local DNN with the stacked denoising auto-encoder (SDAE) as the complete MTC-AE scheme to yield lower classification error caused by corrupted inputs and reach even better performance. The structure of MTC-AE is shown in Figure \ref{fig:model}.

Experiments on benchmark dataset IEMOCAP show that our method outperforms the existing state-of-the-art methods, achieving $64.8\%$ unweighted accuracy with leave-one-speaker-out (LOSO) 10-fold cross-validation. 

\section{Multi-channel Auto-Encoder}
\label{sec:format}
\begin{figure}[t]
\centering
\includegraphics[width=0.9\columnwidth]{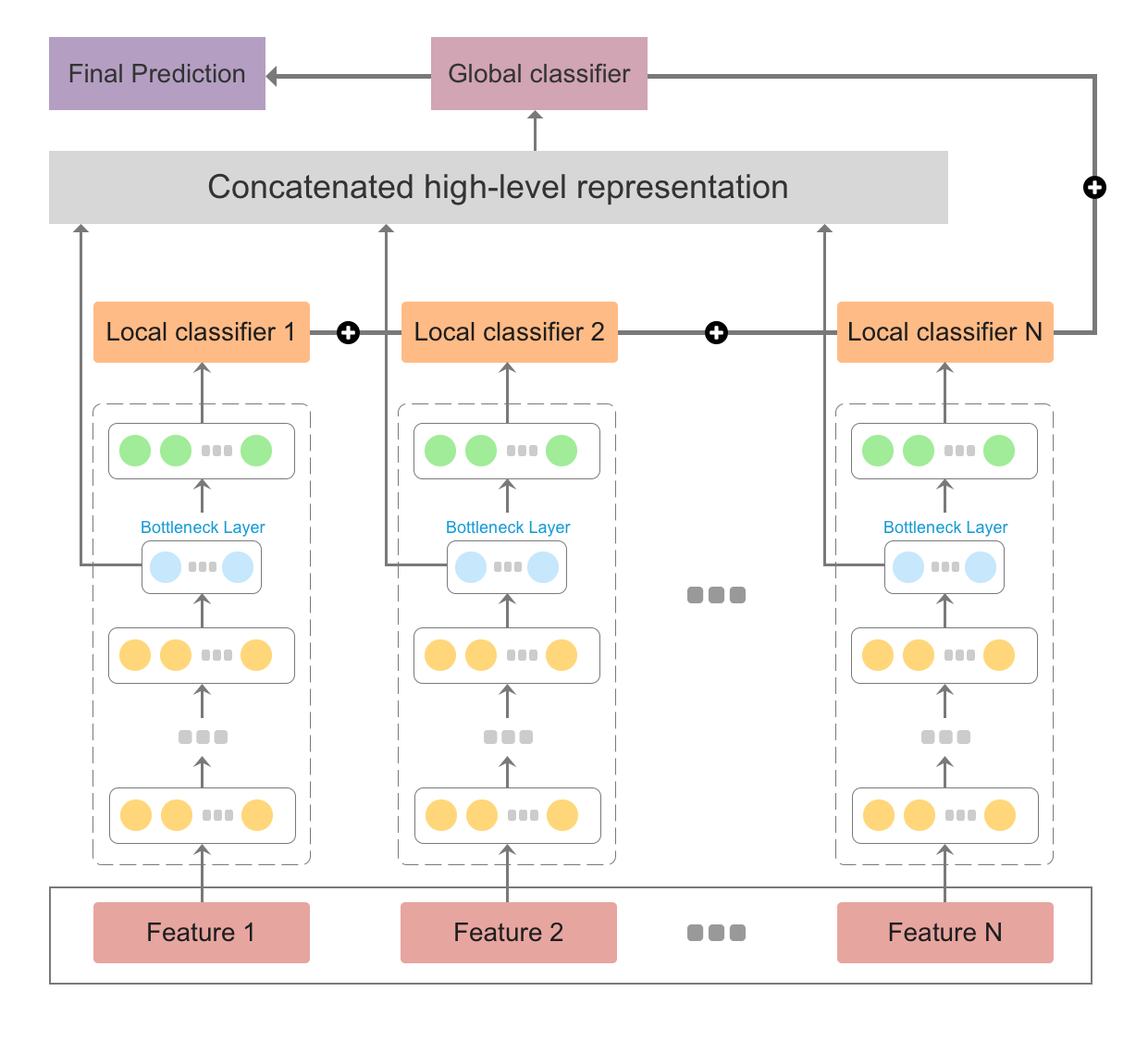}
\caption{The structure of Multi-channel Auto-encoder(MTC-AE). The yellow layers in each local calssifier are pre-trained by SDAEs. The blue layers are bottleneck layers which are concatenated for global classifier. And the blue layers are fully connected layer with random initialization for each local classifiers. All the predictions are fused for final prediction.}
\label{fig:model}
\end{figure}

In this section, we will interpret the complete scheme of MTC-AE in detail. In each local DNN, we use SDAEs for initialization before supervised training, to denoise corrupted versions of the inputs. Local classifiers are obtained after training from each local DNN based on each LLD. Meanwhile, we take bottleneck layers of each local DNN and concatenate them together as the total representation to train a global classifier. Training global classifier and the local ones simultaneously does not only take both relevance and independence of different LLDs into consideration, but also relieve overfitting caused by high-dimensional input and amount-limited training data.  Figure \ref{fig:model} shows the detailed structure.

\subsection{Local Initialization}

Because of the large amount of local classifiers, our network structure is relatively complex. Initialization of network parameters is quite a key factor affecting performance. 
In voice data, it is usual that data gets corrupted. Corruption of data effectively influence the performance of training.
Inspired by DBNF\cite{Gehring2013Extracting}, we utilize SDAEs for initialization. As shown in Figure \ref{fig:model}, the lowest two layers in our local DNN are pre-trained by using SDAEs in an unsupervised manner. 
A denoising auto-encoder operates mostly like a traditional auto-encoder, except that denoising auto-encoder takes the corrupted version $\widetilde{x}$ as its input, which is generated by a corruption process $q(\widetilde{x}\mid x)$ operated on the clean data $x$, and is meant to reconstruct the original data $x$. Generally, a random fraction of the elements of $x$ are set to be zero as the corrupted version $\widetilde{x}$. The reconstruction process can be written mathematically:
\begin{equation}
\begin{aligned}
    \hat{x} = \sigma_{2}(W_{2}\cdot\sigma_{1}(W_{1}\cdot\widetilde{x}+b_{1})+b_{2})
\end{aligned}
\end{equation}
where $\widetilde{x}$ is the input, $\hat{x}$ is the output, $W_{1}, W_{2}$ are the linear transforms, $b_{1}, b_{2}$ are the biases, and $\sigma_{1}, \sigma_{2}$ are the non-linear active functions. In our formulation, we use the ELU function \cite{Clevert2015Fast} for both $\sigma_{1}, \sigma_{2}$:
\begin{equation}
\begin{aligned}
    \sigma_{1}(x) = \sigma_{2}(x)=
    \begin{cases}
    x  & x\geq{0}\\
    \alpha\cdot (e^{x}-1)& x<0
\end{cases}
\end{aligned}
\end{equation}
where $\alpha$ is set to be $1$ in our paper. We use the back-propagation algorithm to train the auto-decoder by minimizing the cost function as following:
\begin{equation}
\begin{aligned}
    \min_{W_1, b_1, W_2, b_2}\sum_{x}{\mid\mid x-\hat{x}\mid\mid ^{2}}+\beta\cdot(\mid\mid W_{1}\mid\mid ^2 + \mid\mid W_{2}\mid\mid ^2)
\end{aligned}
\end{equation}
where $\beta$ is the hyper parameter to limit the influence of regularization term. In our experiments, $\beta$ is set to be $10^{-4}$. 

After the previous auto-encoder is trained, the hidden representation is regarded as the original data for training, and conveyed to the next auto-encoder. Totally, for each local classifier, two auto-encoders are trained for initialize the lowest two layers.

\subsection{Joint Fine-tuning}

After training SDAEs, a bottleneck layer, a hidden layer and a classification layer will be connected to form a feed-forward neural network for each local classifier. The three layers attached on the top are initialized randomly, while the previous layers are initialized with the auto-encoder weights. Then, as shown in Figure  \ref{fig:model}, the feed-forward neural network is utilized as a block to form a local classifier in our framework.

It's important to note that only measuring the local classifiers does consider information of independence sufficiently, but ignores the relevance among different features on the other hand. Therefore, we concatenate the bottleneck layer of each local classifier as the global representation to train the global classifier, which is initialized randomly. The global classifier takes the relevance of each local representation into consideration, and measures it effectively. 

Moreover, in order to optimize the whole system considering both relevance and the independence, we use a single objective function to train local classifiers and the global one simultaneously:
\begin{equation}
\begin{aligned}
\label{equ:cost}
    \min_{\phi_{g}, \phi_{l,i}} \lambda\cdot\mathcal{H}(p(x), q_{\phi_{g}}(x))+ (1-\lambda)\cdot \sum_{i=1}^{N}{ \mathcal{H}(p(x), q_{\phi_{l,i}}(x))}
\end{aligned}
\end{equation}
where $p(\cdot)$ is the true distribution of one-hot label and $q(\cdot)$ is the approximating distribution. $\phi_{g}$ is the parameters of global classifier and $\phi_{l,i}$ is the parameters of $i$th classifier. $N$ is the number of local classifiers and $\lambda$ is the weight coefficient that between $0$ and $1$, in our experiments, we set $\lambda$ to be $0.1$.
Specially, for $\lambda=0$, only the global classifiers are included in the framework, and for $\lambda=1$, local classifiers are included instead of the "global classifiers".
$\mathcal{H}(\cdot)$ is a function that returns cross-entropy between an approximating distribution and a true distribution that can be written mathematically:
\begin{equation}
\begin{aligned}
    \mathcal{H}(p(x), q(x))=-\sum_{x}{p(x)\cdot\log (q(x))}
\end{aligned}
\end{equation}
For the objective function, Eq.\ref{equ:cost}, can be optimized based on back-propagation algorithm.

Since our model contains many classifiers, we fuse the outputs of each classifier by summation simply with a weight parameter $\gamma$ between global classifier and local classifier,  which is set to be $0.95$ in this paper:
\begin{equation}
\begin{aligned}
    \mathcal{F}(x)=\gamma\cdot q_{\phi_{g}}(x)+ (1-\gamma)\cdot \sum_{n=1}^{N}{q_{\phi_{l,i}}(x)}
\end{aligned}
\end{equation}
The maximum output position of $\mathcal{F}(x)$ is regarded as the final prediction result.

\section{Results and Discussion}
\label{sec:experiments}
\subsection{Dataset}

The IEMOCAP \cite{busso2008iemocap} database contains approximately 12 hours' audio-visual conversations of 10 speakers in English, with them manually segmented into utterances. The database contains the following categorical labels: anger, happiness, sadness, neutral, excitement, frustration, fear, surprise, and others. In our experiment, we form a four-class emotion classification dataset containing \{happy, angry, sad and neutral\} after merging happiness and excitement categories as the happy category only, to compare with the former state-of-the-art methods as mentioned in section \ref{sub:experimental_details}. Table \ref{tab:IEMOCAP_data} presents the detail utterance number and the corresponding percentage of each category.

\begin{table}[h]
\centering 
\begin{tabular}{c|c c c c|c}
\hline
Category&Happy&Anger&Sad&Neutral&Total\\
\hline 
Utterances & 1636  & 1103 & 1084 & 1708 & 5531\\ 
\hline
Percentage & 29.6  & 19.9 & 19.6 & 30.9 & 100\\ 
\hline
\end{tabular}
\caption{The number of utterances and their corresponding percentage for each emotion category.}
\label{tab:IEMOCAP_data}
\end{table}
\subsection{Experimental Details}
\label{sub:experimental_details}

\textbf{Evaluation} 
We performed all evaluations using 10-fold LOSO cross-validation, to stay in the same manner as most approaches, so that there is no speaker overlap between the training and test data. As for the method to evaluating the performance, we utilize the unweighted accuracy (UA), which have been used in several previous emotion challenges. UA is quite a good measurement in this case since the class distribution is imbalanced.

\textbf{Feature Extraction}
We utilize openSMILE toolkit \cite{Eyben2010Opensmile} to extract statistics features which was used in the INTERSPEECH 2010 Paralinguistic Challenge \cite{Schuller2010The}, as discussed in \cite{xia2017multi}. Totally, 1582-dimensional features are generated by extracting 38 kinds of LLDs shown in Table \ref{tab:IEMOCAP_feature}, and applying 21 statistics functions shown in Table \ref{tab:IEMOCAP_statictics}. Details of these features can be found in \cite{Schuller2010The}. 

\begin{table}[h]
\centering 
\begin{tabular}{c}
\hline
Low level Descriptors (LLDs)\\
\hline 
PCM loudness\\
MFCC [0-14]\\
log Mel Frequency Band [0-7]\\
Line Spectral Pairs (LSP) Frequency [0-7]\\
F0 by sub-harmonic summation\\
F0 Envelope\\
Voicing probability\\
Jitter local\\
Jitter difference of difference of periods (DDP)\\
Shimmer local\\
\hline
\end{tabular}
\caption{38-dimensional frame-level acoustic features}
\label{tab:IEMOCAP_feature}
\end{table}

\begin{table}[h]
\centering 
\begin{tabular}{c}
\hline
Statistics Functions\\
\hline 
Position maximum/minimum\\
Arithmetic mean, standard deviation\\
Linear regression coefficients 1/2\\
Linear regression error quadratic/absolute\\
Quartile 1/2/3\\
Quartile range 2-1/3-2/3-1\\
Percemtile 1/99\\
Percemtile range 99-1\\
Up-level time 75/99\\
\hline
\end{tabular}
\caption{21 kinds of statistics functions applied on LLDs}
\label{tab:IEMOCAP_statictics}
\end{table}

\begin{table*}[t]
\centering
\begin{tabular}{c c|c|c}
\hline
Method &Reference & Validation Setting &UA  (\%) \cr
\hline
Ensemble of SVM Trees &[APSIPA ASC, 2012] &10-fold LOSO &60.9\\

Replicated Softmax Models + SVM &[ISCAS, 2014] &10-fold LOSO &57.4\\

CNN Feature + MKL Classifier&[ICDM, 2016]&10-fold  &61.3\\

Contextual LSTM &[ACL, 2017]&First 8 speakers for training &57.1\\

Attention-based RNN&[ICASSP, 2017]&4 Sessions for training &58.8\\

Deep Multi-layered Neural Network&[Neural Networks, 2017]& 8-fold LOSO &60.9\\

Multi-task DBN Feature + SVM&[Trans-AC, 2017]&10-fold LOSO &\textbf{62.4}\\

\hline
MTC-DNN&Our Method &10-fold LOSO &62.7\\

MTC-AE&Our Method &10-fold LOSO &\textbf{64.8}\\

\hline
\end{tabular}
\caption{The performance on IEMOCAP dataset with different models and comparison with the state of the art based on unweighted accuracy}
\label{tab:result}
\end{table*}

\textbf{Network Setting}
In our experiments, 38 local classifiers consist in the total framework, corresponding 38-dimensional frame-level acoustic features mentioned above.

For per-training the SDAE, Adam\cite{Kingma2014Adam} is used for optimization with $0.0003$ learning rate for $200$ iteration, and the batch size is $64$. Input vectors are corrupted by applying masking noise to set a random $20\%$ of their elements to zero. Each auto-encoder contains $400$ hidden units and the next one is trained on top of it when finish training.

For fine-tuning process, a bottleneck layer with $30$ units is added, and followed by a new hidden layer with $100$ units. For global classifier, the new hidden layer is set with $1000$ units.
Similarly, Adam is used for optimization with $0.0003$ learning rate for $1000$ iteration, and batch size is $64$. After each epoch, the current model was evaluated on validation set, and the model performing best on this set is used for testing.

All of these processes are done on GPUs using Keras toolkit.

\textbf{State-of-the-art Methods}
To evaluate the effectiveness of proposed framework, we compare the performance of emotion classification with some state-of-the-art methods based on IEMOCAP as following:

[APSIPA ASC, 2012] \cite{rozgic2012ensemble} proposed an ensemble of trees of binary SVM classifiers to address the sentence-level multimodal emotion recognition problem. 

[ISCAS, 2014] \cite{shah2014multi} proposed a multi-modal framework for emotion recognition
using bag-of-words features and undirected, replicated softmax topic models.

[ICDM, 2016] \cite{poria2016convolutional} feed features extracted by deep convolutional neural networks(CNN) into a multiple kernel learning classifier to do multimodal emotion recognition.

[ACL, 2017] \cite{poria2017context} This paper propose a LSTM-based model to capture contextual information between utterance-level features in the same video.

[ICASSP, 2017] \cite{mirsamadi2017automatic} This paper study automatically discovering emotionally relevant speech features using a deep recurrent neural network(RNN) and a local attention base feature pooling strategy.

[Neural Networks, 2017] \cite{Fayek2017Evaluating} proposed speech emotion recognition system to empirically explore feed-forward and recurrent neural network architectures and their variants.

[Trans-AC, 2017] \cite{xia2017multi} This paper propose a framework for acoustic emotion recognition based on the deep belief network (DBN) framework.

\subsection{Experimental Results and Analysis}
Table \ref{tab:result} compares the classification performance of different models from different state-of-the-art literatures based on IEMOCAP dataset. Our experiments was run with 10-fold LOSO as most of the previous work did, and was carried out separately based on whether to use SDAEs to pre-train the network. We call the network without SDAEs as MTC-DNN. The highest accuracy on unweighted accuracy was $62.4\%$ by using muti-task deep belief network to obtain feature representation, and utilizing SVM as the classifier \cite{xia2017multi} before our work. Comparing to this, the two methods we proposed, named MTC-DNN and MTC-AE, improved the UA by $0.3\%$ and $2.4\%$, respectively.
Both the two methods outperformed the existing state-of-the-art methods. It well proves that taking the independence of different features into consideration does help to improve the performance of the system. It also adds a strong regularization to the whole system, and makes the system more discriminating for classification. And it is remarkable to point out that using SDAEs to pre-train the network helps to further improve the classification accuracy. SDAEs explicitly reduces the existing noises from the data set, and helps to accelerate in convergence of the network.

\section{Conclusion}
In this paper, we proposed a novel architectures named  MTC-AE for speech emotion recognition. Multiple local DNNs not only keeps the information from the independence of each feature, but also adds a regularization to the whole scheme, helping to relieve over fitting. Moreover, bottleneck layers from local DNNs are concatenated altogether to form a strong classifier. SDAEs is utilized  to initialize complex networks. Experiments show that our method significantly outperforms the existing state-of-the-art methods with UA on IEMOCAP dataset.

\vfill\pagebreak
\bibliographystyle{IEEEbib}
\bibliographystyle{plain}

\end{document}